\documentclass[10pt,preprintnumbers,twocolumn,nofootinbib,floatfix,superscriptaddress]{revtex4}
\usepackage{epsfig}
\usepackage{amssymb}
\usepackage{amsmath}
\usepackage{mathrsfs}

\makeindex \textheight=9.25in \textwidth=6.5in \topmargin=-0.25in
\begin{document}
\title{Geometrically Engineerable Chiral Matter in M-Theory}
\author{Jacob L. Bourjaily}
%\email{jbourj@sns.ias.edu}
\affiliation{Joseph Henry Laboratories, Princeton University, Princeton, NJ 08544}
%\affiliation{School of Natural Sciences, Institute for Advanced Study, Princeton, NJ 08540}??
\author{Sam Espahbodi}
%\email{espahbod@umich.edu}
\affiliation{Physics Department, University of Michigan, Ann Arbor, MI 48109}
%\author{James D. Wells}
%\email{jwells@umich.edu}
%\affiliation{Physics Department, University of Michigan, Ann Arbor, MI 48109}
%\affiliation{CERN, Theory Division (PH-TH), CH-1211 Geneva 23, Switzerland}
\date{8$^{\mathrm{th}}$ April 2008}

\begin{abstract}
We present a classification of the massless chiral matter representations that can arise locally in M-theory on $G_2$ through geometrically engineered singularities. We will find that several of the more exceptional singularities could have applications for model building.% Because some of these geometries give rise to chiral fields in reducible, complex representations, some care must be taken to determine which specific representations result. This point was raised by Berglund and Brandhuber in \cite{Berglund:2002hw}, and we will settle it here through a slight refinement of the original discussion of Acharya and Witten \cite{Acharya:2001gy}.
\end{abstract}

\maketitle
%%%%%%%%%%%%%%%%%%%%%%%%%%%%%%%%%%%%%%%%%%%DRAFT COMMAND%%%%%%%%%%%%%%%%%%%%%%%%%%%%%%%%%%%%%%%%%%%
%\draft
%%%%%%%%%%%%%%%%%%%%%%%%%%%%%%%%%%%%%%%%%%%DRAFT COMMAND%%%%%%%%%%%%%%%%%%%%%%%%%%%%%%%%%%%%%%%%%%%
%\vspace{-0.35cm}
\section{Introduction}\vspace{-0.35cm}
Any phenomenologically viable string vacuum must have non-abelian gauge symmetries with massless charged matter. These features are of course manifest in heterotic string models, and can be readily built into type II string theory by the addition of \mbox{D-branes} and orientifold planes. Both of these frameworks have duals in M-theory where non-abelian gauge symmetries and massless charged matter arise through geometrically-engineered singularities within the compactification manifold.

The local geometric structures giving rise to massless matter in M-theory, resolutions of $ADE$-orbifolds, are well understood and are themselves trivially classifiable. However, there has been some uncertainty in the literature about which matter representations result from the more exceptional of these singularities \cite{Berglund:2002hw}. We will revisit the argument given by Acharya and Witten in \cite{Acharya:2001gy}, and resolve any potential ambiguities.

An explicit tabulation of the representations that result from each conical singularity is useful not only because the group theory data required for this goes beyond that found in standard references (e.g. Slansky's review \cite{Slansky:1981yr}), but also because it may lead to new ideas in string phenomenology. 

It may seem that such a classification would have few implications for phenomenology. Indeed, perhaps the first phenomenological questions asked of any string model would be the multiplicity of different matter representations and how disparate representations are coupled in the superpotential. These questions are usually understood as topological, and therefore particularly problematic in M-theory where no compact $G_2$-manifold with massless matter is known concretely---even though many are expected to exist by duality to string theory. 

One way to avoid this problem in M-theory would be to consider the more exceptional, non-compact geometries that give rise to multiple massless matter representations, coupled together in a calculable way; so long as the matter produced by such a local patch is anomaly-free, any effects involving the entire compact manifold may be viewed as sub-leading---effectively excising our ignorance of quantum gravity from phenomenology. This philosophy has already led to some very interesting results in type IIb string theory \cite{Verlinde:2005jr,Buican:2006sn}. The classification of locally engineerable matter representations in M-theory gives us a starting point for applying those ideas here. 

To give an idea of the somewhat unusual structure that can be found in M-theory, consider the conical singularity \mbox{`$E_7\to SO_{10}\times SU_2$'} listed in Table \ref{exceptional_res}. This geometry gives rise to massless matter transforming in the representation \vspace{-0.2cm}\begin{equation}(\mathbf{16},\mathbf{2})\oplus(\mathbf{10},\mathbf{1})\label{so10_su2}\vspace{-0.2cm}\end{equation} of $SO_{10}\times SU_2$ gauge theory---two full families of the Standard Model together with a single pair of Higgs doublets. Because the matter content is anomaly free\footnote{Every geometry we consider will have a $U_1$-anomaly; but these $U_1$'s are not necessarily dynamical, and their anomalies can be cancelled by the inflow mechanism \mbox{\cite{Witten:2001uq,Atiyah:2001qf}}.}, it is not unreasonable to consider it by itself in the manifold; but we would also be free to suppose that another $\mathbf{16}$ of $SO_{10}$ is generated elsewhere in the manifold. Because of the proximity of the Higgs fields to two of the families, we would find a vast hierarchy of yukawa couplings separating these two from the third generation.

This qualitative discussion of yukawa couplings can be made very concrete if the grand-unified gauge groups of these local models are broken through geometric unfolding as described in \mbox{\cite{Bourjaily:2007kv,Bourjaily:2007vw,Bourjaily:2007vx}}. The unfolded local geometry has the structure necessary to generate a leading superpotential that is in principle calculable over its entire moduli space. In this way, it may be possible to construct a local geometry in M-theory that includes all three families of the Standard Model coupled in a phenomenologically viable way, while still remaining agnostic about quantum gravity---i.e. the global topology of the compactification manifold. A complete discussion would go well beyond the scope of this paper. 

This paper is organized as follows. In section \ref{geometrical_engineering} we review the basic building blocks used to geometrically engineer charged matter in M-theory, and in section \ref{classification} we present a classification of the representations that can arise locally. 
%\newpage
\vspace{-0.35cm}
\section{Geometrical Engineering in M-Theory}\label{geometrical_engineering}\vspace{-0.35cm}
In this section we briefly review the geometric structures giving rise to massless charged matter in M-theory. This is meant only as a pointillistic sketch, sufficient to understand the basis for the classification presented in section \ref{classification}, and to address the potential ambiguities raised in \cite{Berglund:2002hw}. The interested reader should also consult the original papers on geometrical engineering in M-theory \cite{Katz:1996xe,Acharya:1998pm,Atiyah:2001qf,Witten:2001uq,Acharya:2001gy}, or the excellent review \cite{Acharya:2004qe}. 

\vspace{-0.35cm}
\subsection{Basic Building Blocks}\vspace{-0.25cm}
Non-abelian gauge symmetry is known to arise in M-theory through the existence of co-dimension four $ADE$-orbifold singularities in the $G_2$-compactification manifold. These singularities span three-cycles in the $G_2$, and are named according to the gauge group that results from compactification, which can be either \mbox{$SU_n(\equiv A_{n-1})$}, \mbox{$SO_{2n}(\equiv D_n)$}, or \mbox{$E_n(\equiv E_n)$}\footnote{We assume that the singularities undergo no outer-automorphisms around the three-cycle. But this point is moot for the local geometries in which we are interested.}. %Add some of the historical citations---Klein, Aspinwall, early Witten, ...? 

Massless charged matter arises locally from co-dimension seven singularities within the compactification manifold. In order for matter to be charged under a particular non-abelian gauge group $H$, it must come from a conical singularity that lies along the singularity $H$---that is, from an isolated point along $H$ where the type is singularity is `worsened.' 

An $ADE$-orbifold singularity of type $H$ can only be worsened into one of higher rank, say $G$; this happens through the shrinking of an additional two-cyle that intersects $H$. As described below, this means that $H$ must come from the moduli space of resolutions of a $G$-type singularity, and so this geometry is called `resolution $G\to H$'. The group-theoretic translation of this geometric criterion is that the resolution $G\to H$ exists if and only if the Dynkin diagram of $H$ can be obtained by removing one node from that of $G$. This is a minor point, but it explains why, for example, there is no resolution \mbox{$E_7\to SU_4\times SU_4$}.

It is worth mentioning that because three-cycles in a seven-manifold do not generically intersect, multiply-charged matter is highly non-generic in M-theory. This fact could help make room for sequestered sectors of supersymmetry breaking, a feature exploited in \cite{Acharya:2006ia,Acharya:2007rc}, and could be a hint that we should seek a more geometrically unified origin of the Standard Model's \mbox{$SU_3\times SU_2$} gauge group, such as the $SU_5$ construction explored in \cite{Witten:2001bf,Friedmann:2002ty}.

\vspace{-0.55cm}
\subsubsection{$ADE$-Orbifolds and Their Resolutions}\vspace{-0.35cm}
Any rank-$n$ gauge group can be completely broken to a product of $n$ $U_1$-gauge symmetries through adjoint Higgs fields. In M-theory, $U_1$-gauge bosons arise from Kaluza-Klein reduction of the supergravity three-form on two-cycles, and so one may guess that $ADE$-singularities are resolvable into a non-singular collection of two-cycles. Because asymptotically away from the singularity, the metric must continue to respect the orbifold structure of $\mathbb{R}^4/\Gamma_{ADE}$, the resolutions of $ADE$-singularities are known as asymptotically locally Euclidean (ALE) spaces. They are non-compact hyper-K\"ahler four-manifolds.

The smooth ALE spaces that resolve $SU_n$-singularities are nothing other than the Gibbons-Hawking gravitational multi-instantons \cite{Gibbons:1979zt}. Hitchin's construction of these spaces in \cite{Hitchin:1994zr} naturally suggested the existence of ALE resolutions for any $ADE$-group, a conjecture that was proven by Kronheimer in \cite{Kronheimer:1989zs}. Kronheimer achieved this by constructing the ALE resolution of each $ADE$-singularity as the Higgs branch of the vacuum manifold of an $\mathscr{N}=2$ quiver gauge theory---the quiver being the Dynkin diagram of the $ADE$-group labeled with Dynkin indices. We will not have need to describe this construction in any real detail; a good treatment, emphasizing its use in M-theory, can be found in \cite{Acharya:2004qe} and \cite{Berglund:2002hw}; and the quiver gauge theories involved are thoroughly analyzed in \cite{Lindstrom:1999pz,Park:1999hu,Albertsson:2001jq}.

A detailed understanding of Kronheimer's construction is not required to understand the geometry of these spaces. The resolution of a rank-$n$ $ADE$-singularity is found to contain $n$ independent two-cycles which are `physically arranged' according to the singularity's Dynkin diagram. By this we mean that a basis of two cycles can be chosen so that their intersection matrix is the negative of the Cartan matrix of the corresponding gauge group \cite{Kronheimer:1989zs}---nodes representing the basis two-cycles, and lines connecting two-cycles that intersect. 

The moduli space of these ALE-spaces is very simple. It can be given completely in terms of the volumes of each of the $n$ two-cycles as measured with respect to each of the three K\"ahler forms. Notice that the three K\"ahler forms naturally give rise to this generalized, three-dimensional notion of the volume of a two-cycle. 

\vspace{-0.35cm}
\subsubsection{Conical Enhancements of $ADE$-Singularities}\vspace{-0.35cm}
Consider the moduli space of resolutions of a $G$-type singularity of rank-$n$. As described above, this space consists of the $n$ three-vectors controlling the `volumes' of each of the $n$ independent two-cycles. Let \mbox{$t\in\mathbb{R}^3$} parameterize the volume of one of the two-cycles in $G$, keeping the rest shrunk; and let $\hat{G}(t)$ denote the partial resolution of $G$ for fixed $t$. For \mbox{$t\neq0$} the geometry of $\hat{G}(t)$ contains one non-vanishing two-cycle together with $(n-1)$-shrunk two-cycles arranged according to a subgroup \mbox{$H\subset G$}. Thus, M-theory on $\hat{G}(t\neq0)$ would give rise to \mbox{$H\times U_1$} gauge theory. 

If we use $\hat{G}(t)$ to define a fibration of resolutions of $G$ over $\mathbb{R}^3$, the resulting non-compact seven-manifold would have a co-dimension four singularity of type $H$, enhanced to one of type $G$ at the origin. This is the concrete meaning of the resolution \mbox{$G\to H$}. 

The physics that results from M-theory compactified on this space can be rigorously studied through its duality with the heterotic string. We will not have need to re-tell that story here; an interested reader should consult \cite{Acharya:2004qe} and the references therein. The important lesson is that this duality allows us to infer the existence of a $G_2$-holonomy metric for $G\to H$, and to determine the representations of massless charged matter that arise in M-theory.

\subsection{Massless Chiral Matter in M-Theory}\vspace{-0.35cm}
In \cite{Acharya:2001gy}, Acharya and Witten determined the representations of massless chiral matter resulting from M-theory compactified on almost any resolution $G\to H$ by solving the Dirac equation explicitly in the dual heterotic context. Only a slight refinement of their conclusion will be necessary to extend it to all resolutions, addressing the questions raised by Berglund and Brandhuber in \cite{Berglund:2002hw}. Once this refinement is made, we will be able to directly state the complete classification of engineerable representations.

In order to highlight where ambiguity arises for some resolutions in M-theory, let us consider the analogous problem of geometrically engineering $\mathscr{N}=2$ hypermultiplets in type IIa string theory. As described by Katz and Vafa \cite{Katz:1996xe}, the localized massless hypermultiplet resulting from the resolution $G\to H$ is given by the $U_1$-charged components of the branching of the adjoint of $G$ into $H\times U_1$. In general, although perhaps not uniquely, we may write this branching as \begin{equation}\mathrm{Adj}(G)=\mathrm{Adj}(H)_0\oplus\mathbf{1}_0\oplus\mathbf{R}\oplus\overline{\mathbf{R}};\label{generic_branching}\end{equation} and so the resolution $G\to H$ will give rise to a massless hypermultiplet with components transforming as $\mathbf{R}\oplus\overline{\mathbf{R}}.$

Now, Acharya and Witten's computation was specifically done for the case when the representation $\mathbf{R}$ in equation (\ref{generic_branching}) is irreducible. They showed that the net number of chiral zero modes was one, meaning that, fixing chirality, either $\mathbf{R}$ or $\overline{\mathbf{R}}$ was a normalizable zero mode, but not both. Notice that when $\mathbf{R}$ is irreducible, this same result follows heuristically from the fact that M-theory on $G\to H$ has only $\mathscr{N}=1$ supersymmetry: the chiral matter must be one half of the hypermultiplet that would have resulted in the analogous $\mathscr{N}=2$ construction, and there is no confusion about how \mbox{$\mathbf{R}\oplus\overline{\mathbf{R}}$} is decomposed into chiral components when $\mathbf{R}$ is an irreducible representation. 

This heuristic argument can be extended to allow one to unambiguously determine the massless chiral matter arising in M-theory from \mbox{$G\to H$} whenever the representation $\mathbf{R}$ appearing in (\ref{generic_branching}) has at most one complex irreducible component. But often this is not the case, and so the argument must be refined. 

As an example, let us consider the resolution \mbox{$E_7\to SU_4\times SU_3\times SU_2\times U_1$} discussed in \cite{Berglund:2002hw}. For this geometry, the $U_1$-charged part of (\ref{generic_branching}) is given by
 \begin{align}&\phantom{\oplus~}(\mathbf{4},\overline{\mathbf{3}},\mathbf{2})_1\oplus(\mathbf{6},\overline{\mathbf{3}},\mathbf{1})_{-2}\oplus(\overline{\mathbf{4}},\mathbf{1},\mathbf{2})_3\oplus(\mathbf{1},\mathbf{3},\mathbf{1})_{-4}\nonumber\\&\oplus(\overline{\mathbf{4}},\mathbf{3},\overline{\mathbf{2}})_{-1}\oplus(\mathbf{6},\mathbf{3},\mathbf{1})_2\oplus(\mathbf{4},\mathbf{1},\overline{\mathbf{2}})_{-3}\oplus(\mathbf{1},\overline{\mathbf{3}},\mathbf{1})_4.\label{big_ex}\end{align}
There are eight inequivalent ways to write this as a representation plus its conjugate: writing $\mathbf{r}_{q}$ for the component of (\ref{big_ex}) with $U_1$-charge $q$, and choosing to associate $\mathbf{r}_{+1}$ with $\mathbf{R}$ rather than $\overline{\mathbf{R}}$, the choices correspond to \begin{equation}\mathbf{R}_{\overrightarrow{\sigma}}=\mathbf{r}_{\phantom{{\sigma}_{1}}\!\!\!\!\!\!+1}\oplus\mathbf{r}_{\sigma_12}\oplus\mathbf{r}_{\sigma_23}\oplus\mathbf{r}_{\sigma_34}\quad\mathrm{with}\quad\sigma_i=(\pm).\end{equation}
Berglund and Brandhuber guessed that the chiral matter coming from this geometry in M-theory would be in the representation $\mathbf{R}_{--+}$, but they saw no reason why this choice was forced upon them, and hoped a more detailed analysis would settle this uncertainty \cite{Berglund:2002hw}.

On general grounds, unless every choice of $\mathbf{R}_{\overrightarrow{\sigma}}$ is possible to engineer in M-theory, there must be a reason why one choice is distinguished. And we would hope that if such a choice exists, it can be made canonically for any resolution \mbox{$G\to H$}. Does there exist any canonical way of writing the $U_1$-charged components of (\ref{generic_branching}) as $\mathbf{R}\oplus\overline{\mathbf{R}}$? There does: regardless of the reducibility of the representations, we can always write (\ref{generic_branching}) unambiguously as \begin{equation}\mathrm{Adj}(G)=\mathrm{Adj}(H)_0\oplus\mathbf{1}_0\oplus\mathbf{R}_+\oplus\overline{\mathbf{R}}_-,\label{correct_split}\end{equation} where the subscript refers to the sign of the $U_1$-charges. This suggests that, up to chirality, the resolution \mbox{$E_7\to SU_4\times SU_3\times SU_2\times U_1$} gives rise to matter transforming in the representation $\mathbf{R}_{+++}$, or, \begin{equation}(\mathbf{4},\overline{\mathbf{3}},\mathbf{2})_1\oplus(\mathbf{6},\mathbf{3},\mathbf{1})_2\oplus(\overline{\mathbf{4}},\mathbf{1},\mathbf{2})_3\oplus(\mathbf{1},\overline{\mathbf{3}},\mathbf{1})_4.\end{equation}

To see that M-theory does indeed make this choice of representations, we should revisit the original discussion of Acharya and Witten in \cite{Acharya:2001gy}. They determined the representation that results from $G\to H$ by solving the Dirac equation for the geometry's heterotic dual. When $\mathbf{R}$ is irreducible, its $U_1$-charge can be set to $+1$ without loss of generality, and this was used to simplify their argument. But their conclusion, that for a representation with positive $U_1$-charge, zero modes of the Dirac operator have fixed chirality (determined by the orientation of the fibration), holds generally\footnote{To see this explicitly, one should follow the argument in \cite{Acharya:2001gy}, replacing the single Dirac operator acting on a mode with charge $q=+1$ with a ``vector'' of Dirac operators indexed by the $U_1$-charges appearing in the branching (\ref{generic_branching}). After this, the analysis of massless modes goes through unchanged for each component separately.}: all the modes appearing in $\mathbf{R}_+$ of (\ref{correct_split}) will have the same chirality. 

This answers the question raised by \cite{Berglund:2002hw}, and allows us to proceed directly to a listing of all the representations that result in M-theory from geometrically engineered singularities.

\section{Classification of Engineerable Representations in M-theory}\label{classification}\vspace{-0.35cm}
We may summarize the discussion above as follows. Massless matter charged under the gauge group $H$ results from isolated places where the $H$-type singularity is enhanced into one of type $G$, written \mbox{$G\to H$}; the groups $G$ into which $H$ can be enhanced are only those obtained by connecting one additional node to the Dynkin diagram of $H$. Up to an overall chirality, the massless matter arising from M-theory on $G\to H$ transforms according to the positive $U_1$-charged components of the decomposition of the adjoint of $G$ into $H\times U_1$. 

And so we now come to the math problem in group theory and representation theory to list every possible resolution \mbox{$G\to H$} and determine the representation that results from each. Although this computation can be worked out with the aid of known methods \cite{Kim:1982jg,Fuchs:1997jv}, the calculation does require representation theory data beyond that which is tabulated in common references (e.g. \cite{Slansky:1981yr}).

For the resolutions of $SU_n$ and $SO_{2n}$, the resulting representations are listed in Table \ref{simple_res}. Resolutions of $E_n$ singularities are listed in Table \ref{exceptional_res}.

\section{Discussion}\vspace{-0.35cm}
One important consequence of the classification presented in Tables \ref{simple_res} and \ref{exceptional_res} is the difficulty of finding a simple three-generation model arising locally in M-theory. Perhaps the most simple geometry would have been $E_8\to E_6\times SU_2$. This was studied in the context of geometrically engineered $\mathscr{N}=2$ models in type IIa in \cite{Bourjaily:2007vx}. But Table \ref{exceptional_res} makes it clear that this geometry in M-theory does not give three families, but two families and a mirror. We should refrain from saying that this would be `one net family' because two conjugate fields can only be projected out through a mass term in the superpotential if there exists a (supersymmetric) three-cycle in the manifold supporting the two singularities. In general, we would expect that most of the chiral matter will remain light. This does not save $E_8\to E_6\times SU_2$ from phenomenological impotency, however, because a theory with a right-handed third generation is almost certainly incompatible with phenomenology.

Several resolutions give rise to two or more families, but none gives an anomaly-free embedding of three generations alone. Perhaps the closest is $E_8\to SO_{10}\times SU_{3}$, which would give three families, each with their own pair of Higgs doublets, together with a single exotic mirror family. The matter content from this singularity has an $SU_3$ anomaly, and so there would necessarily be other Standard-Model singlet fields, transforming under the $SU_3$-flavour symmetry. But of course, it would be very hard for a model with a mirror fourth family to be phenomenologically viable (but see e.g. \cite{Kribs:2007nz}).

As described in the introduction, one way to get three families alone would be to patch together anomaly-free collections of mutually isolated conical singularities. In this way, it may be possible to find a viable three-family model. Such a construction would naturally have large hierarchies among yukawa couplings---potentially explaining the large hierarchies we observe in the Standard Model.

\begin{table}[t]\caption{Resolutions of $SU_n$ and $SO_{2n}$.\label{simple_res}}
\begin{tabular}{lcll}
\toprule
\multicolumn{3}{c}{Resolution}&Representation\\
\hline 
$SU_{n+1}$&$\longrightarrow$&$SU_n$&$\mathbf{n}$\\
$SU_{n+m}$&$\longrightarrow$&$SU_n\!\times\!SU_m$&$(\mathbf{n},\overline{\mathbf{m}})$\\
$SO_{2n}$&$\longrightarrow$&$SU_n$&$\left[\mathbf{n}\otimes\mathbf{n}\right]_{a}$\\
$SO_{2(n+1)}$&$\longrightarrow$&$SO_{2n}$&$\mathbf{2n}$\\
$SO_{2(n\hspace{-0.02cm}+\hspace{-0.02cm}m)}\footnote{For $m\geq2$, with the understanding that $SO_4\simeq SU_2\times SU_2$.}$&$\longrightarrow$&$SU_n\times SO_{2m}\times U_1$&$(\mathbf{n},\mathbf{2m})_1$\\&&&$\oplus(\left[\mathbf{n}\otimes\mathbf{n}\right]_{a},\mathbf{1})_2$\\
\botrule
\end{tabular}\end{table}

\begin{table*}[t]\caption{Exceptional Resolutions.\label{exceptional_res}}
\begin{tabular}{rllc}
\toprule
\multicolumn{2}{c}{Resolution}&Representation\\
\hline
$E_6\longrightarrow$&$SO_{10}\times U_1$&$\mathbf{16}_1$\\
$\phantom{E_6}\longrightarrow$&$SU_5\times SU_2\times U_1$&$(\mathbf{10},\mathbf{2})_1\oplus(\overline{\mathbf{5}},\mathbf{1})_2$\\
$\phantom{E_6}\longrightarrow$&$SU_3\times SU_3\times SU_2\times U_1\hspace{0.4cm}$&$(\overline{\mathbf{3}},\mathbf{3},\mathbf{2})_1\oplus(\mathbf{3},\overline{\mathbf{3}},\mathbf{1})_2\oplus(\mathbf{1},\mathbf{1},\mathbf{2})_3$\\
$\phantom{E_6}\longrightarrow$&$SU_6\times U_1$&$\mathbf{20}_1\oplus\mathbf{1}_2$\\\hline
$E_7\longrightarrow$&$E_6\times U_1$&$\mathbf{27}_1$\\
$\phantom{E_7}\longrightarrow$&$SO_{10}\times SU_2\times U_1$&$(\mathbf{16},\mathbf{2})_1\oplus(\mathbf{10},\mathbf{	1})_2$\\
$\phantom{E_7}\longrightarrow$&$SU_5\times SU_3\times U_1$&$(\mathbf{10},\overline{\mathbf{3}})_1\oplus(\overline{\mathbf{5}},\mathbf{3})_2\oplus(\mathbf{5},\mathbf{1})_{3}$\\
$\phantom{E_7}\longrightarrow$&$SU_4\times SU_3\times SU_2\times U_1$&$(\mathbf{4},\overline{\mathbf{3}},\mathbf{2})_1\oplus(\mathbf{6},\mathbf{3},\mathbf{1})_2\oplus(\overline{\mathbf{4}},\mathbf{1},\mathbf{2})_3\oplus(\mathbf{1},\overline{\mathbf{3}},\mathbf{1})_4$\\
$\phantom{E_7}\longrightarrow$&$SU_6\times SU_2\times U_1$&$(\mathbf{15},\mathbf{2})_1\oplus(\overline{\mathbf{15}},\mathbf{1})_2\oplus(\mathbf{1},\mathbf{2})_3$\\
$\phantom{E_7}\longrightarrow$&$SO_{12}\times U_1$&$\mathbf{32}_1\footnote{This is the $\mathbf{32}'$ in the notation of \mbox{reference \cite{Slansky:1981yr}}.}\oplus\mathbf{1}_2$\\
$\phantom{E_7}\longrightarrow$&$SU_{7}\times U_1$&$\overline{\mathbf{35}}_1\oplus\mathbf{7}_2$\\\hline
$E_8\longrightarrow$&$E_7\times U_1$&$\mathbf{56}_1\oplus\mathbf{1}_2$\\
$\phantom{E_8}\longrightarrow$&$E_6\times SU_2\times U_1$&$(\mathbf{27},\mathbf{2})_1\oplus(\overline{\mathbf{27}},\mathbf{1})_2\oplus(\mathbf{1},\mathbf{2})_3$\\
$\phantom{E_8}\longrightarrow$&$SO_{10}\times SU_3\times U_1$&$(\mathbf{16},\mathbf{3})_1\oplus(\mathbf{10},\overline{\mathbf{3}})_2\oplus(\overline{\mathbf{16}},\mathbf{1})_3\oplus(\mathbf{1},\mathbf{3})_4$\\
$\phantom{E_8}\longrightarrow$&$SU_5\times SU_4\times U_1$&$(\mathbf{10},\mathbf{4})_1\oplus(\overline{\mathbf{5}},\mathbf{6})_2\oplus(\mathbf{5},\overline{\mathbf{4}})_3\oplus(\overline{\mathbf{10}},\mathbf{1})_4\oplus(\mathbf{1},\mathbf{4})_5$\\
$\phantom{E_8}\longrightarrow$&$SU_5\times SU_3\times SU_2\times U_1$&$(\mathbf{5},\overline{\mathbf{3}},\mathbf{2})_1\oplus(\mathbf{10},\mathbf{3},\mathbf{1})_2\oplus(\overline{\mathbf{10}},\mathbf{1},\mathbf{2})_3\oplus(\overline{\mathbf{5}},\overline{\mathbf{3}},\mathbf{1})_4\oplus(\mathbf{1},\mathbf{3},\mathbf{2})_5\oplus(\mathbf{5},\mathbf{1},\mathbf{1})_6$\\
$\phantom{E_8}\longrightarrow$&$SU_7\times SU_2\times U_1$&$(\overline{\mathbf{21}},\mathbf{2})_1\oplus(\mathbf{35},\mathbf{1})_2\oplus(\mathbf{7},\mathbf{2})_3\oplus(\overline{\mathbf{7}},\mathbf{1})_{4}$\\
$\phantom{E_8}\longrightarrow$&$SO_{14}\times U_1$&$\mathbf{64}_1\oplus\mathbf{14}_2$\\
$\phantom{E_8}\longrightarrow$&$SU_{8}\times U_1$&$\mathbf{56}_1\oplus\overline{\mathbf{28}}_2\oplus\mathbf{8}_3$\\
\botrule
\end{tabular}
\end{table*}

\section{Acknowledgements}\vspace{-0.35cm}
It is a pleasure to thank Edward Witten, Bobby Acharya, Sergei Gukov, Nima Arkani-Hamed, and Paul Langacker for many helpful discussions. This work started in discussions with James Wells, who  offered many helpful suggestions on early drafts of this paper. JLB was supported in part by a Graduate Research Fellowship from the National Science Foundation, and through the hospitality of the Michigan Center for Theoretical Physics, and the Simons Workshop on Mathematical Physics 2007.

\newpage

%\bibliographystyle{hieeetr}
%\bibliographystyle{h-physrev}
%\bibliography{geo_engin_reps}

%\end{document}

\end{document}